\documentclass[prl,twocolumn,superscriptaddress,floatfix,preprintnumbers,amssymb,amsmath]{revtex4}
\usepackage{graphicx}
\usepackage{dcolumn}
\usepackage{bm}
\usepackage[latin1]{inputenc}
\usepackage[mathscr]{eucal}
\usepackage{epsfig}
\usepackage{rotating}
\usepackage{multirow}

\preprint{ }

\begin{document}
\title{Wide-band detection of the third moment of shot noise by a hysteretic Josephson junction}

\author{A.V. Timofeev}
\affiliation{Low Temperature Laboratory, Helsinki University of
Technology, P.O. Box 3500, 02015 TKK, Finland}
\affiliation{Institute of Solid State Physics, Russian Academy of
Sciences, Chernogolovka, 142432 Russia}
\author{M. Meschke}
\affiliation{Low Temperature Laboratory, Helsinki University of
Technology, P.O. Box 3500, 02015 TKK, Finland}
\author{J.T. Peltonen}
\affiliation{Low Temperature Laboratory, Helsinki University of
Technology, P.O. Box 3500, 02015 TKK, Finland}
\author{T.T. Heikkil\"a}
\affiliation{Low Temperature Laboratory, Helsinki University of
Technology, P.O. Box 3500, 02015 TKK, Finland}
\author{J.P. Pekola}
\affiliation{Low Temperature Laboratory, Helsinki University of
Technology, P.O. Box 3500, 02015 TKK, Finland}


\begin{abstract}

We use a hysteretic Josephson junction as an on-chip detector of the
third moment of shot noise of a tunnel junction. The detectable
bandwidth is determined by the plasma frequency of the detector,
which is about 50 GHz in the present experiment. The third moment of
shot noise results in a measurable change of the switching rate when
reversing polarity of the current through the noise source. We
analyze the observed asymmetry assuming adiabatic response of the
detector.

\end{abstract}
\maketitle Studies of shot noise in mesoscopic conductors are
presently of great interest, revealing detailed information on
microscopic mechanisms of electronic transport
\cite{blanter00,nazarov03,glattli04}. The theory of full counting
statistics (FCS) of electrons \cite{levitov96} determines the
probability distribution of current fluctuations and its n$th$ order
moments. Experimentally, however, measurements of higher moments
beyond the variance remain difficult especially in the interesting
high frequency regime, mainly because of weak signals and demanding
filtering requirements. The first experiments on the third moment of
shot noise of a tunnel junction were performed on samples connected
remotely to the detector \cite{reulet03,bomze05}. Yet the most
natural way to investigate noise in nano-structures would seem to be
via an on-chip detector. In recent measurements of Gustavsson et al.
\cite{gustavsson06} and of Fujisawa et al. \cite{fujisawa06}, single
electrons tunneling through quantum dots could be counted directly.
The noise correlations were observed up to the third order over a
bandwidth of a few tens of kHz. The latest remarkable results on
observation of the fourth and the fifth moments of current through a
quantum dot \cite{gustavsson0607} demonstrated on-chip detection of
FCS in the low frequency regime. As regards to high-frequency
on-chip detectors of non-Gaussian noise, a Josephson junction (JJ)
appears to be a very attractive one according to various theoretical
predictions \cite{JJdet,ankerhold06,peltonen06,sukhorukov06}. On the
experimental side, the influence of non-Gaussian noise on the
conductance of a Coulomb blockaded JJ has been measured in
\cite{lindell04}.
\begin{figure}
\begin{center}
\includegraphics[width=0.475
\textwidth]{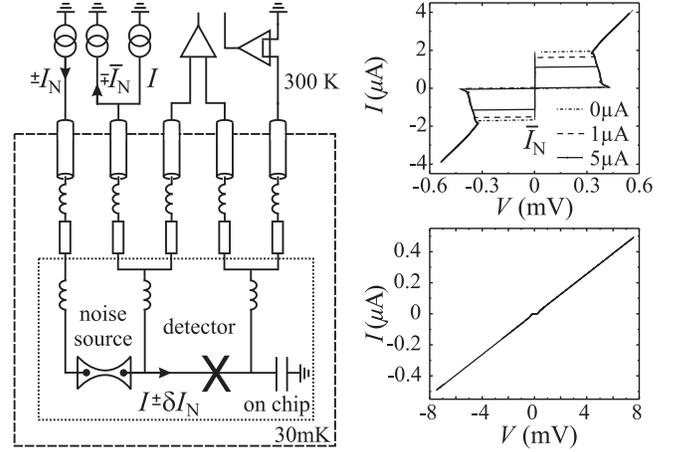} \caption{Measurement scheme for on-chip
detection of noise. A Josephson junction detects the current
fluctuations arising from the noise source. Only heavily filtered DC
lines (through cold resistors, line inductances and thermocoax
cables) connect the setup with the room temperature amplifiers and
current sources. On top right we show $IV$ characteristics of the
detector junction taken with different values of average current
$\bar{I}_{\rm N}$ through the NIS noise source. The $IV$ of this NIS
source is depicted in the lower right graph at small
currents.}\label{fig1}
\end{center}
\end{figure}
In this Letter we present an experimental observation and a
theoretical interpretation of the influence of wide-band third order
fluctuations on escape of a hysteretic JJ threshold detector from
its zero voltage state.

Our analysis of the experiment is based on the following model. Let
us assume for a moment that the Josephson potential of the detector
is strictly harmonic with oscillation (plasma) frequency
$\omega_{\rm p} =\sqrt{\frac{2eI_{\rm C}}{\hbar C}}[1-(I/I_{\rm
C})^{2}]^{1/4}$, where $I_{\rm C}$ and $C$ are the critical current
and capacitance of the junction at bias current $I$, respectively.
Then, on the level of the second moment, noise affects the escape of
a JJ predominantly by promoting resonant activation at $\omega_{\rm
p}$ such that shot noise induced transitions between adjacent energy
levels $j$ and $j-1$ in the potential are given by $\gamma_{j,j-1} =
\frac{j}{2\hbar \omega_{\rm p} C}S_I(-\omega_{\rm p})$ (excitation)
and $\gamma_{j-1,j} = \frac{j}{2\hbar \omega_{\rm p}
C}S_I(+\omega_{\rm p})$ (relaxation), where $S_I(\omega)$ is the
spectral density of noise at (angular) frequency $\omega$. These
rates lead to dynamics in the potential with effective temperature
\cite{jp05}
\begin{equation} \label{Tstar}
T^* \simeq \hbar \omega_{\rm p}/[2k_{\rm B}{\rm
arcoth}(1+\frac{QF|\bar{I}_{\rm N}|}{2I_{\rm C}(1-(I/I_{\rm
C})^2)^{1/2}})].
\end{equation}
Here we have assumed that the bath temperature is $T \ll
\hbar\omega_{\rm p}/k_{\rm B}$, i.e., the detector is in macroscopic
quantum tunneling (MQT) regime in the absence of shot noise, and
that the voltage $V$ across the scatterer is large enough such that
we are in pure shot noise regime: $|eV| \gg k_{\rm B}T,
\hbar\omega_{\rm p}$. In Eq. \eqref{Tstar} $Q$ is the quality factor
of the junction {\sl at plasma frequency}, and $\bar{I}_{\rm N}$ is
the average value of the noise current $I_{\rm N}$ through the
scatterer junction with Fano factor $F$ \cite{fano}. Unlike in the
earlier experiments by some of us \cite{jp05}, in the present
measurements we can determine $Q$ independently by detecting the
crossover from escape dynamics to under-damped phase diffusion (UPD)
\cite{UPD} as described in the experimental part below. This leaves
no free parameters to determine the consistency between the measured
escape rate
\begin{equation} \label{Gamma}
\Gamma \simeq \frac{\omega_{\rm p}}{2\pi}\exp(-\Delta U/k_{\rm
B}T^*)
\end{equation}
and $T^*$ given by independent measurements of the parameters in Eq.
\eqref{Tstar}. In \eqref{Gamma} $\Delta U \simeq
\frac{4\sqrt{2}}{3}E_{\rm J}(1-I/I_{\rm C})^{3/2}$ is the barrier
height of the potential well with Josephson energy $E_{\rm J} \equiv
\hbar I_{\rm C}/2e$.

The linear coupling to current fluctuations $\delta I$ in the
Josephson potential is of the form $-\frac{\hbar}{2e}\delta I
\hat{\varphi}$, where $\hat{\varphi}$ is the phase operator. It is
proportional to the sum of the creation and annihilation operators
of the harmonic oscillator, and therefore, in contrast to the
influence of the second moment, the third order fluctuations do not
induce transitions between adjacent levels in a harmonic well
\cite{heikkila06,peltonen06}. Therefore, up to the third order, the
influence of resonant activation in a harmonic potential is
exclusively that due to the second order fluctuations. There are
weak corrections to the results above, if we allow the potential to
be anharmonic. Yet in the case of the third order fluctuations such
corrections vanish unless the third order spectral densities are
assumed to have frequency dependence \cite{heikkila06,peltonen06}.
Therefore we assume in what follows that the third order effects
come from non-resonant fluctuations only, at sub-plasma frequencies.
Throughout we ignore the effects related to higher than third order
fluctuations. Under these conditions, in analogy to adiabatic
Gaussian noise \cite{martinis88}, the contribution of the third
moment to the escape rate from the well is determined by the
instantaneous value of the fluctuating bias current. The asymmetry
$\Delta \Gamma/\Gamma_{\rm ave}\equiv (\langle\Gamma^+\rangle -
\langle\Gamma^-\rangle)/[(\langle\Gamma^+\rangle +
\langle\Gamma^-\rangle)/2]$ between escape rates
$\langle\Gamma^\pm\rangle$, averaged over the adiabatic
fluctuations, at different polarities of either pulse or noise
currents can then be written as \cite{peltonen06}
$\Delta \Gamma/\Gamma_{\rm ave} \simeq -\frac{1}{3}(\frac{\partial
B}{\partial I})^3 \langle \delta I_{\rm N}^3\rangle$. Here $B$ is
the exponent in the expression of the tunneling rate and $\langle
\delta I_{\rm N}^3\rangle$ is the third moment of current
fluctuations at the detector.
For thermal activation of Eq. \eqref{Gamma}, with $B \equiv \Delta
U/k_{\rm B}T^*$, one then obtains
\begin{equation} \label{asymm1}
\Delta \Gamma/\Gamma_{\rm ave} \simeq
\frac{16\sqrt{2}}{3}(\frac{\hbar}{2e})^3 (1-I/I_{\rm C})^{3/2}
\frac{\langle \delta I_{\rm N}^3\rangle}{(k_{\rm B}T^*)^3}.
\end{equation}
Here, $\langle \delta I_{\rm N}^3\rangle \equiv (\frac{\Delta
\omega}{2\pi})^2S_3$, where $\Delta \omega \sim \omega_{\rm p}$ is
the bandwidth of adiabatic fluctuations, and $S_3 =
F_3e^2\bar{I}_{\rm N}$ is the low frequency limit of the third order
spectral density with Fano factor $F_3$.

The measurement scheme is shown in Fig.~\ref{fig1}. The detector is
an Al/AlOx/Al JJ with an area of $\sim2$ $\mu$m$^2$ and it stays
initially in a superconducting state. Another tunnel junction, the
noise source, is biased with a current $\pm I_{\rm N}$, such that it
is driven far from equilibrium into the shot noise regime. All
samples were fabricated by electron beam lithography and shadow
evaporation and they were measured via filtered signal lines at a
bath temperature of $\simeq 30$ mK. In order to detect only the
fluctuations of $I_{\rm N}$, the balancing current $\mp \bar I_{\rm
N}$ of opposite polarity is applied such that no DC-component due to
$I_{\rm N}$ passes through the detector. Owing to the small
magnitude of the third order fluctuations, careful DC-balancing is
extremely important. The residual non-balanced current across the
detector is measured repeatedly by a low input impedance current
amplifier. However, the resulting correction never exceeded 1 nA
during the measurements.
The on-chip inductors in Fig. \ref{fig1} are long (few mm) and
narrow (2 $\mu$m) superconducting lines. In all the samples, the
detector was connected to a large contact pad on the chip (few
mm$^2$ in area), which served as a capacitive short to ground at
high frequencies. To probe the fluctuations of $I_{\rm N}$, we
applied trapezoidal current pulses of height $I$ through the
detector with $\Delta t=$ 100 $\mu$s - 1 ms duration, and with a 1
ms delay between two pulses. Typically $10^3 - 10^4$ pulses at each
value of $I$ were repeated, and the escape probability $P$ was
obtained as the fraction of pulses leading to an escape from the
supercurrent state. When increasing the average value of $I_{\rm
N}$, shot noise enhances the escape rate, leading to suppression of
the escape threshold current \cite{jp05}. To extract information
about the third moment, we applied four different combinations of
current polarities, which we call $I^{++}$, $I^{--}$, $I^{+-}$, and
$I^{-+}$, where the superscripts refer to the signs of $I$ and
$I_{\rm N}$, respectively. The first two combinations should lead to
identical escape characteristics, which should differ, in the
presence of odd moments, from the last two identical combinations.
We normally measure the separation of the histograms, $I^+ - I^-$,
corresponding to equal escape probability $P$ in the two pulse/noise
current directions. By a simple geometric argument, assuming linear
dependence between escape probability and current over a short
interval (weak third order effects), we obtain
\begin{equation} \label{asymmcurr}
I^+ - I^- \simeq (1-P) \ln (1-P) (\frac{\partial I}{\partial P})
\Delta \Gamma/\Gamma_{\rm ave}.
\end{equation}

\begin{table}
\caption{Parameters of the samples.} \label{table}
\begin{tabular}{|c|ccc|cc|}

\hline
 \multirow{2}{*}{Sample} & \multicolumn{3}{c}{detector} \vline & \multicolumn{2}{c}{noise junction} \vline \\
 & $I_{\rm C}$ ($\mu$A)  & $C$ (fF)  & $\omega_{\rm p}/2\pi$ (GHz) & $R_{\rm N}$ (k$\Omega$)  & $C_{\rm N}$ (fF)  \\
\hline
NIS & 2.88 & 80 & 53 &15.4   & 8 \\
SIS & 2.51 & 80 & 49 &1.9 & 6 \\
REF & 2.36 & 75 & 49 & 0 & - \\
\hline
\end{tabular}
\end{table}

\begin{figure}
\begin{center}
\includegraphics[width=0.5
\textwidth]{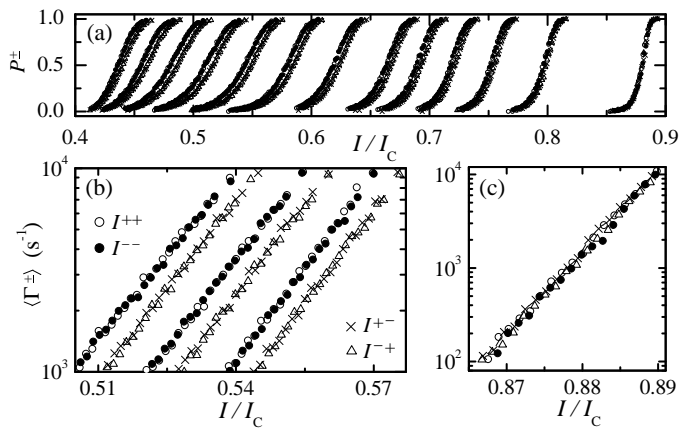} \caption{Escape characteristics of Sample NIS.
(a) Histograms taken with the four different polarity configurations
of noise and pulse currents, and at several values of noise current:
$|\bar{I}_{\rm N}|= 0...3.8$ $\mu$A from right to left. (b) Escape
rates corresponding to $|\bar{I}_{\rm N}|= 2.2,2.4$, and $2.6$
$\mu$A from right to left displaying the splitting of data with
$I^{++}$ and $I^{--}$ currents from those with $I^{+-}$ and $I^{-+}$
currents, respectively. (c) Escape rates with $\bar{I}_{\rm N}= 0$
in MQT regime.}\label{fig2}
\end{center}
\end{figure}

We present data of three samples, see Table I. Sample NIS had a
normal metal-insulator-superconductor Cu/AlOx/Al noise source. In
Sample SIS the noise source was another Al/AlOx/Al JJ, but with a
much smaller critical current than in the detector. In the reference
sample, Sample REF, the noise source was replaced by a
superconducting line. In Table I $R_{\rm N}$ denotes the normal
state resistance of the noise junction, and $\omega_{\rm p}$ is
taken at $I=0$.

Here we describe data of Sample NIS unless otherwise specified.
Figure \ref{fig1} shows current-voltage characteristics of the
detector and the noise source junction. The detector $IV$s have been
measured at different values of noise current. Besides the standard
hysteretic character and suppression of switching threshold upon
increasing $\bar{I}_{\rm N}$, there are two important features to
observe here. (i) The $IV$s are vertical in the supercurrent branch,
such that $V\le 0.2 $ $\mu$V up to the switching current. This means
that the rate of phase diffusion events from one well to another is
$f_{2\pi} =2eV/h \le 100$ MHz. (ii) The gap voltage is the same at
all values of noise current up to 5 $\mu$A; this is higher than all
those currents used for further analysis of data. Thus there is no
excess overheating of the bath and the only relevant temperature is
then $T^*$. Figure \ref{fig2} shows escape characteristics with
rates $\langle\Gamma^\pm\rangle$ that are related to the measured
escape probabilities $P^\pm$ via
$P^\pm=1-e^{-\langle\Gamma^\pm\rangle \Delta t}$. Here the current
pulses were $\Delta t = 480$ $\mu$s long. The measurements were
performed at various values of $\bar{I}_{\rm N}$ using the four
polarity configurations. Suppression of the mean bias current and
change in the width with increasing $\bar{I}_{\rm N}$ is evident
from the histograms in Fig \ref{fig2} (a). Moreover, as to the
different polarity configurations, the $I^{++}$ and $I^{--}$ data
indeed lie together for the same magnitude of $\bar{I}_{\rm N}$, and
the suppression of switching threshold for this pair is higher than
for $I^{+-}$ and $I^{-+}$. The latter two are again grouped
together, as depicted in Fig. \ref{fig2} (b). This shift is the
manifestation of the skewness of the current distribution, where the
sign of the third moment is positive. The values of $I_{\rm C}$ and
$C$ of the detector in Table I are obtained from fits to MQT escape
\cite{weiss}; Fig.~\ref{fig2} (c) shows the rates. The obtained
value of $I_{\rm C}$ agrees with the Ambegaokar-Baratoff prediction
\cite{ambegaokar62}, and $C$ with that based on the size of the
junction.

\begin{figure}
\begin{center}
\includegraphics[width=0.47
\textwidth]{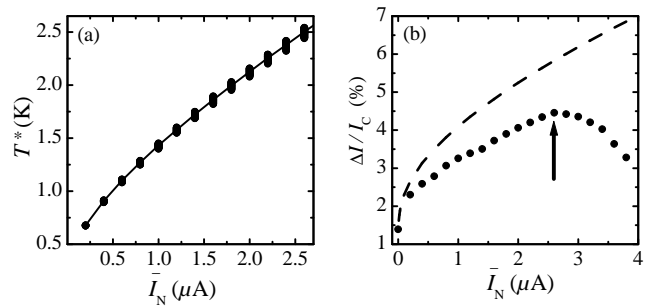} \caption{Results related to the second moment
of shot noise. (a) Effective temperature extracted from measured
escape histograms using Eq. \eqref{Gamma}. At each noise current
$T^*$ has been evaluated at various values of escape probability in
the range 0.1 - 0.9. The solid line is obtained using Eq.
\eqref{Tstar} with $F=1$ and $Q=2.5$ determined from the bias
current at which the underdamped phase diffusion sets in. This is
shown by the arrow in (b), where width of the histogram has been
plotted against $\bar{I}_{\rm N}$. The dashed line is from the
effective temperature model, ignoring phase diffusion.}\label{fig3}
\end{center}
\end{figure}

In Fig. \ref{fig3} we have analyzed the escape histograms of Fig.
\ref{fig2} based on Eq. \eqref{Gamma} to extract effective
temperature at various levels of $\bar{I}_{\rm N}$. We obtain in (a)
a set of values of $T^*$, spread within about $\pm$2\% around the
mean, for each $\bar{I}_{\rm N}$ at different values of bias current
$I$ along a histogram. For clarity, we show data of one pulse
configuration ($I^{+-}$) only, because the third order effects are
weak: the variation in $T^*$ between different configurations is
only about 1\% at maximum. Figure \ref{fig3} (b) shows the measured
width, defined as $\Delta I$ between $P=0.9$ and $P=0.1$ of the same
escape histograms. We see an increase up to about $\bar{I}_{\rm N} =
2.6$ $\mu$A, in fair agreement with the thermal activation model
(dashed line). Above this value of $\bar{I}_{\rm N}$ the width
starts to decrease. We attribute the maximum to the crossover from
escape dynamics to UPD \cite{UPD}, now induced by the increase of
{\sl effective} temperature and subsequent decrease of the switching
threshold current. From the value of $I/I_{\rm C} \simeq 0.52$ for
the $P=0.5$ switching threshold at this noise current, we obtain the
quality factor $Q \equiv \frac{4I_{\rm C}}{\pi I} \simeq 2.5$.
(Similar procedure gave $Q=2.5$ also for Sample SIS.) Based on this
value of $Q$ and the values of $I_{\rm C}$ and $C$ from independent
measurements, we are then able to draw the theoretical line based on
Eq. \eqref{Tstar} in Fig. \ref{fig3} (a).
\begin{figure}
\begin{center}
\includegraphics[width=0.47
\textwidth]{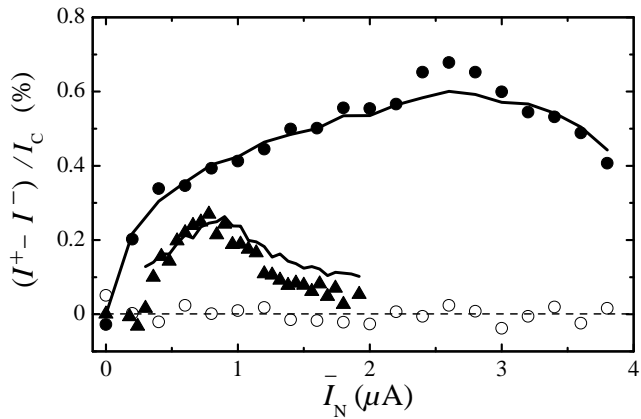} \caption{Difference in switching currents of
the threshold detector under reversal of the relative polarities of
bias and noise currents. Data of Sample NIS are shown by filled
dots, of Sample SIS by triangles and data of the reference sample,
Sample REF, by open dots. The solid lines are the result of the
theoretical model for Samples NIS and SIS. At currents $\bar{I}_{\rm
N}\le 0.2$ $\mu$A the noise source of Sample SIS is in
superconducting state, and the signal due to the third moment
vanishes.}\label{fig4}
\end{center}
\end{figure}

Figure \ref{fig4} shows the influence of the odd moments on escape
threshold of all the three samples. The quantity $I^+ - I^-$ has
been extracted as $\frac{1}{2}[(I^{+-}+I^{-+})-(I^{++}+I^{--})]$,
all taken at $P=0.5$. We see that the signal due to the third
moment, plotted against $\bar{I}_{\rm N}$ again, of Sample NIS
(filled dots) has a maximum value of about 0.6\% of $I_{\rm C}$.
This is to be contrasted to the essentially vanishing signal of the
reference sample (open circles). The solid line following closely
the data of Sample NIS is the result of the theoretical model
according to Eqs. \eqref{asymm1} and \eqref{asymmcurr}. The
non-trivial parameters needed to construct this curve, $\partial
I/\partial P$ and $T^*$, have been extracted from experiment. Note
that values of $F$ and $Q$ are not needed here since we measure
$T^*$ directly. In addition, we have set $F_3 = 1$, the value for
Poissonian noise of a tunnel junction, and $\Delta \omega =
\omega_{\rm p}$, which is a very natural choice for our model
including adiabatic frequencies only. We also show similar data of
Sample SIS (triangles). Here, the bandwidth was taken to be $0.8
\omega_{\rm p}$ for the theory line. It may seem surprising at the
first view that the results follow the theoretical prediction also
in the regime where the detector is in the UPD regime ($\bar{I}_{\rm
N} > 2.6$ $\mu$A for Sample NIS). Yet this is quite natural, because
the frequency of the phase diffusion events is $<100$ MHz, whereas
the relevant bandwidth of noise affecting the phase particle extends
up to $\sim 50$ GHz. Therefore it suffices to study escape from one
well only.

We have observed the third moment of shot noise up to frequencies of
$\sim50$ GHz, and interpreted this observation by assuming that the
third order effects manifest themselves in our JJ system as an
average thermal escape rate in a potential which varies according to
adiabatically fluctuating bias currents. A similar problem has been
analyzed theoretically in \cite{ankerhold06} using an effective
Fokker-Planck method and very recently in \cite{sukhorukov06} by
generalized stochastic path integral methods. The results of our
model coincide with \cite{ankerhold06} up to numerical prefactor of
the order of unity in the limit of large $Q$, and with
\cite{sukhorukov06} excluding the circuit corrections that are weak
in our case. Finally, although the data presented here can be
interpreted quantitatively, it remains a challenge to engineer the
radio-frequency environment on the chip to the degree that no
uncertainty in the relevant bandwidth would remain. However, it is
already possible to envision experiments where different mesoscopic
noise sources use a common JJ detector of third order fluctuations,
and where a tunnel junction serves as a reference scatterer.

We thank J. Ankerhold, H. Grabert, T. Ojanen, H. Pothier and E.
Sukhorukov for useful discussions and Academy of Finland for
financial support. A.V.T. thanks Finnish Academy of Science and
Letters for a scholarship.


\begin{thebibliography}{99}


\bibitem{blanter00} Y.M. Blanter and M. B\"uttiker, Phys. Rep. {\bf 336}, 1
(2000).

\bibitem{nazarov03} {\sl Quantum Noise in Mesoscopic Physics}, edited by Yu.V
Nazarov (Kluwer, Dordrecht, 2003).

\bibitem{glattli04} {\sl Quantum Information and Decoherence in Nanosystems},
edited by D. C. Glattli, M. Sanquer, and J. Tran Thanh Van. (Proc.
XXXIX Recontres de Moriond, La Thuile, Italy, 2004.)

\bibitem{levitov96} L.S. Levitov, H.W. Lee, and G.B. Lesovik, J. Math. Phys.
{\bf 37}, 4845 (1996).

\bibitem{reulet03} B. Reulet, J. Senzier, and D.E. Prober, Phys. Rev. Lett.
{\bf 91}, 196601 (2003); B. Reulet \emph{et al.}, in Ref.
\cite{nazarov03}.

\bibitem{bomze05} Yu. Bomze \emph{et al.},
Phys. Rev. Lett. {\bf 95}, 176601 (2005).

\bibitem{gustavsson06} S. Gustavsson \emph{et al.}, Phys. Rev. Lett.
{\bf 96}, 076605 (2006).

\bibitem{fujisawa06} T. Fujisawa, T. Hayashi, R. Tomita, Y. Hirayama,
Science {\bf 312}, 1634 (2006).

\bibitem{gustavsson0607} S. Gustavsson \emph{et al.},
cond-mat/0607192.

\bibitem{JJdet} J. Tobiska and Yu.V. Nazarov, Phys. Rev. Lett. {\bf 93},
106801 (2004); J.P. Pekola, ibid. {\bf 93}, 206601 (2004); T.T.
Heikkil\"a, P. Virtanen, G. Johansson, and F.K. Wilhelm, ibid. {\bf
93}, 247005 (2004); E.B. Sonin, Phys. Rev. B 70, 140506(R) (2004);
J. Ankerhold and H. Grabert, Phys. Rev. Lett. {\bf95}, 186601
(2005).

\bibitem{ankerhold06} J. Ankerhold, cond-mat/0607020 (2006).

\bibitem{peltonen06} J.T. Peltonen, A.V. Timofeev, M. Meschke, and J.P.
Pekola, cond-mat/0611593, J. Low Temp. Phys., to be published
(2006).

\bibitem{sukhorukov06} E.V. Sukhorukov and A.N. Jordan,
cond-mat/0611783.

\bibitem{lindell04} R.K. Lindell {\sl et al.}, Phys. Rev. Lett. {\bf 93},
197002 (2004).


\bibitem{jp05} J.P. Pekola {\sl et. al.}, Phys. Rev. Lett. {\bf95}, 197004 (2005).

\bibitem{fano} This factor is ideally the Fano factor of the noise
source \cite{blanter00}, but it does include the frequency
dependence of the surrounding circuit also. In the present
experiment the circuit was designed in such a way that its influence
on $F$ is supposedly weak.

\bibitem{UPD} J.M. Kivioja {\sl et al.}, Phys. Rev. Lett. {\bf
94}, 247002 (2005); V.M. Krasnov {\sl et al.}, ibid. {\bf 95},
157002 (2005); J. M\"annik {\sl et al.}, Phys. Rev. B {\bf 71},
220509(R) (2005).

\bibitem{heikkila06} T.T. Heikkil\"a and T. Ojanen, cond-mat/0609133.

\bibitem{martinis88} J.M. Martinis and H. Grabert, Phys. Rev. B {\bf 38}, 2371 (1988).

\bibitem{weiss} U. Weiss, {\sl Quantum Dissipative Systems}, 2nd edition (World Scientific, Singapore, 1999).

\bibitem{ambegaokar62} V. Ambegaokar and A. Baratoff, Phys. Rev. Lett. {\bf 10}, 000486
(1963).

\end{thebibliography}
\end{document}